\documentclass[%
 reprint,
 amsmath,amssymb,
 aps,
]{revtex4-2}

\usepackage{graphicx}
\usepackage{dcolumn}
\usepackage{bm}
\usepackage{xcolor}

\usepackage{hyperref}
\usepackage{braket}
\usepackage{soul}
\usepackage{siunitx}

\begin{document}
\setstcolor{red}

\preprint{APS/123-QED}

\title{Electrically-Small Rydberg Sensor for Three-Dimensional Determination of rf $k$-Vectors}

\author{Peter K. Elgee}
\email{peter.k.elgee.ctr@army.mil}
\author{Kevin C. Cox}
\author{Joshua C. Hill}
\author{Paul D. Kunz}
\author{David H. Meyer}
\affiliation{%
 DEVCOM Army Research Laboratory, 2800 Powder Mill Rd, Adelphi MD 20783, USA
}%
\date{\today}

\begin{abstract}
    We present an electrically-small Rydberg atom electric field sensor with the ability to extract the three-dimensional $k$-vector of an elliptically polarized radio frequency (rf) field.
    In most mediums, the $k$-vector (or wave vector) provides the direction of propagation of an electromagnetic wave.
    Our method uses a field vector measurement at a single point in space and is thus compatible with a sensor volume that is arbitrarily small compared to the carrier wavelength.
    We measure the $k$-vector of a circularly polarized signal field with average absolute errors in the polar and azimuthal angles of $33$~\unit{\milli\radian} and $43$~\unit{\milli\radian} respectively, and statistical noise of 1.3~\unit{\milli\radian\per\sqrt{\hertz}} and 1.5~\unit{\milli\radian\per\sqrt{\hertz}}.
    Additionally, we characterize the performance of the sensor as a function of the ellipticity of the input field and the size of the sensing region.
    We find that the sensor works over a broad range of ellipticities and validate that an electrically-small sensing region is optimal.
    
\end{abstract}

\maketitle

Electric field sensors are powerful tools for field characterization and communication.
However, measuring the electric field vector at a single point in space is insufficient to fully characterize an electromagnetic wave.
To determine the $k$-vector of an incoming linearly-polarized plane wave, which is defined to be perpendicular to the wavefronts of the wave, one must either make a spatially-extended measurement, make simultaneous measurements of the electric and magnetic field vectors \cite{Kanda84}, or measure the momentum vector of the wave for example through particle recoil \cite{Chu93}.
With electric field measurements alone, only spatially extended measurements that cover an appreciable fraction of the wavelength will provide the $k$-vector, such as with an interferometer \cite{Ho04}, focusing element \cite{Nan11}, or other highly directional antenna, and the signal goes to zero with sensor size \footnote{In theory, a highly directional antenna can be made arbitrarily small \cite{Bouwkamp45}, but such an antenna is generally not realizable in practice.}.
These extended methods are the most common, and work by measuring the phase difference between points in space, either directly or through interference, to deduce the $k$-vector.
This restriction is relaxed for elliptically-polarized plane waves where the electric field vector traces out an ellipse, and the $k$-vector is constrained to be normal to this ellipse.
Thus, for elliptically-polarized plane waves the $k$-vector can be determined up to a sign by measuring the electric field vector at a single point in space.

Rydberg electric field sensors offer compelling capabilities beyond classical antennas, including their large tunability \cite{Downes20, Jau20, Meyer21}, non-absorptive sensing \cite{Holloway14}, SI traceability \cite{Holloway14, Anderson21}, and small size \cite{Fan14}.
However, previous $k$-vector measurement methods with Rydberg electric field sensors have relied on at least two spatially separated measurements to determine the angle of a source \cite{Robinson21, Yan23, Mao24,Rajavardhan25}.
This method degrades as the sensor becomes significantly electrically-small.

In this work, we demonstrate an electrically-small quantum sensor that measures the $k$-vector for elliptically-polarized rf fields.
Building off our previous development of a Rydberg atom polarimeter \cite{Elgee24}, we extract the three-dimensional $k$-vector by measuring the polarization ellipse with a single optical probe measurement.
This method allows the sensor to be arbitrarily small relative to the carrier wavelength, and thus operates down to quasi-DC fields.
We quantitatively assess the k-vector measurement, the impact of interaction region size relative to the rf wavelength, and the dependence of the accuracy and precision of the measurement on the ellipticity of the input field.
The three-dimensional and inherently electrically-small nature of our method opens up new possibilities in sub-wavelength field characterization and alternatives to extended apertures or arrays in radar systems or radio frequency telescopes as in most cases the $k$-vector corresponds to the propagation direction of the wave.

\begin{figure}[b]
    \includegraphics[width = 3.375 in]{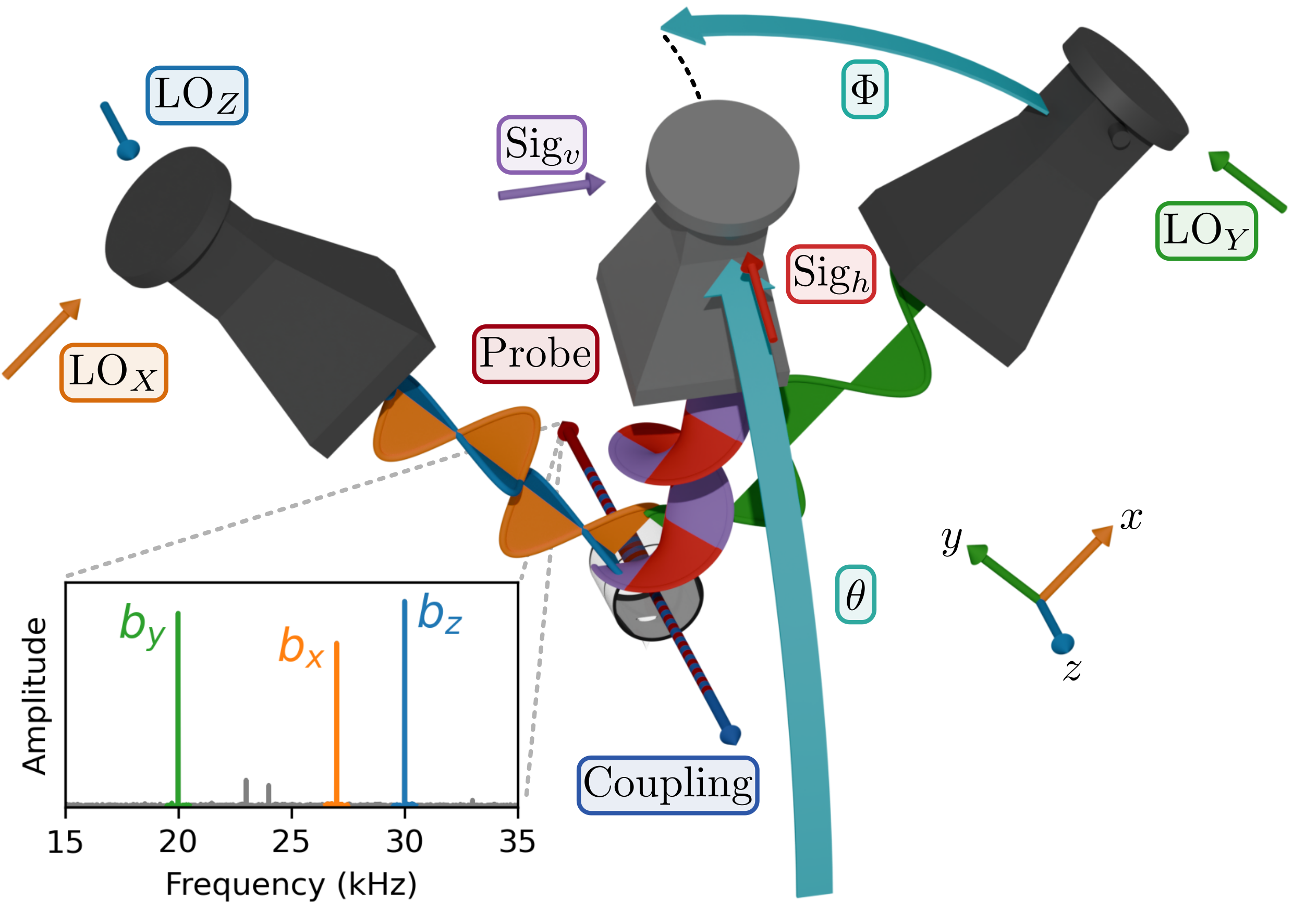}
    \caption{
    A diagram of the sensor.
    This shows the orientation of the LO and signal field inputs and rf polarizations relative to the optical beams and vapor cell.
    The Cartesian coordinate system for the field vectors, and spherical coordinate system for the $k$-vector are defined here.
    An example spectrum extracted from the probe transmission is shown in the inset.
    }
    \label{fig:diagram}
\end{figure}

The basic principle of our sensor is discussed in Ref.~\cite{Elgee24}, and the apparatus is detailed in Fig.~\ref{fig:diagram}.
$^{85}$Rb atoms in a vapor cell are excited to a Rydberg state through a two-photon process with probe and coupling lasers where they become sensitive to a signal rf field (Sig) resonant with an atomic transition between Rydberg states.
We measure the amplitude and phase of the incoming signal field projected along each cardinal axis through rf heterodyne measurements \cite{Simons19, Jing20, Meyer21} with three local oscillators (LO$_X$, LO$_Y$, LO$_Z$) polarized along each axis.
The rf heterodyne beats are present in the Autler-Townes splitting of the Rydberg states, and measured by the absorption of the probe beam through electromagnetically induced transparency (EIT) spectroscopy.
By applying different detunings to each LO we separately extract from the Fourier transform of the probe beam transmission the beat amplitude for each axis $|b_x|$, $|b_y|$, $|b_z|$, and the relative phases $\phi_{xy}$, $\phi_{xz}$ of the field projection along each axis, where $\phi_{ij}$ is the phase difference between the projection along $\hat{i}$ and $\hat{j}$ ($\phi_i - \phi_j$).
These quantities provide a measurement of the electric field vector in three dimensions.

We have modified the previous apparatus \cite{Elgee24} to implement $k$-vector measurements.
Using a gimble, we vary the azimuthal ($\Phi$), and polar ($\theta$) angles of the rf horn antenna that transmits the signal field as shown in Fig.~\ref{fig:diagram}.
The azimuthal angle is set within $\Phi \in [\pi/8, 3\pi/8]$ and the polar angle is set within $\theta \in [\pi/4, 3\pi/4]$.
The signal field is transmitted by a dual polarization horn, allowing us to adjust the ellipticity of the field by changing the relative phases of the two separate inputs (labeled $\text{Sig}_{v}$ and $\text{Sig}_{h}$).
We maintain an rms amplitude of 0.06(1)~\unit{\volt\per\meter} for the signal field, and 0.14(2)~\unit{\volt\per\meter} for the LOs throughout the experiment.
Unless otherwise specified, we use a $2.5$~\unit{\centi\meter} long vapor cell, and a signal field on resonance with the Rydberg transition 70D$_{5/2} \rightarrow 69$F$_{7/2}$ at 6.64~\unit{\giga\hertz} (4.51~\unit{\centi\meter}).
This configuration keeps the interaction region smaller than the rf wavelength, and suppresses the signal averaging effect discussed later.
We also keep LO detunings small ($\delta_X = 27$~\unit{\kilo\hertz}, $\delta_Y = -20$~\unit{\kilo\hertz}, $\delta_Z = 30$~\unit{\kilo\hertz}) to avoid any negative effects from the presence of Earth's magnetic field.
The coordinate axes are defined such that $\hat{z}$ is along the optical axis and the spherical coordinates are conventionally defined relative to our gimbal mechanism (this differs from the coordinate system in Ref.~\cite{Elgee24}).

{\it $k$-vector measurement}---The three beat amplitudes ($|b_x|$, $|b_y|$, $|b_z|$), and two relative phases ($\phi_{xy}$, $\phi_{xz}$) are sufficient to extract the k-vector up to a sign in the following way.
The field vector in time is reconstructed (up to an absolute phase) as

\begin{equation}\label{eq:field_reconstruction}
\begin{split}
    \vec{E}(t) \propto \ &|b_x|\cos(\omega_0 t)\hat{x}\\
    + &|b_y|\cos(\omega_0 t - \phi_{xy}) \hat{y}\\
    + &|b_z|\cos(\omega_0 t - \phi_{xz})\hat{z},
\end{split}
\end{equation}
where $\omega_0$ is the signal frequency.
Provided that the signal field is not purely linearly polarized, the field traces out an ellipse in three dimensions, and we can extract two orthogonal vectors along this ellipse when $t = 0$ and $t = \pi/2\omega_0$:
\begin{align}\label{eq:orthogonal_vectors}
        \vec{v_1} &= |b_x|\hat{x} + |b_y|\cos(-\phi_{xy})\hat{y} + |b_z|\cos(-\phi_{xz})\hat{z}\\
    \vec{v_2} &= |b_y|\sin(-\phi_{xy})\hat{y} + |b_z|\sin(-\phi_{xz})\hat{z}.
\end{align}
For a plane wave, the $k$-vector must be orthogonal to the field vector at all times, and we can extract it up to a sign from the cross product of $\vec{v_1}$ and $\vec{v_2}$
\begin{equation}\label{eq:k_vector}
    \hat{k} = \pm \frac{\vec{v_1}\times\vec{v_2}}{|\vec{v_1}||\vec{v_2}|}.
\end{equation}
In this work, we will only be considering the unit $k$-vector, as the frequency of the field is presumed to be known or easily determined from the heterodyne measurements.
\begin{figure*}
    \includegraphics[width = 6.5in]{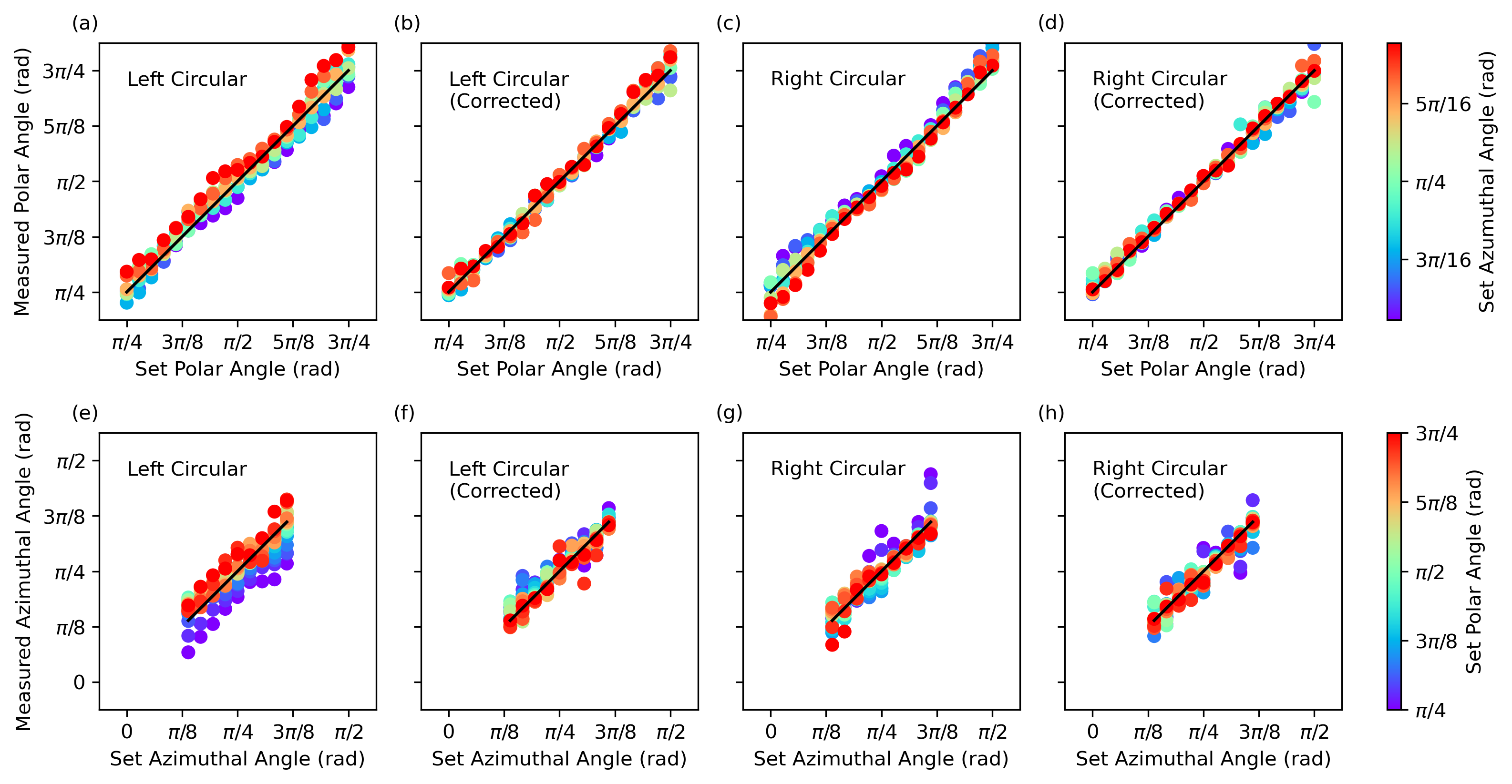}
    \caption{A comparison of the set and measured polar ((a) - (d)), and azimuthal ((e) - (h)) angles of the signal field $k$-vector.
    The results for left circularly polarized light are shown on the left ((a), (b), (e), (f)), and for right circularly polarized light on the right ((c), (d), (g), (h)).
    The data in (a), (c), (e), and (f) have an average calibration applied, as described in the text, which largely accounts for LO amplitudes and phases.
    These data have an average accuracy of 60~\unit{\milli \radian} in $\theta$, and 63~\unit{\milli\radian} in $\Phi$.
    The data in (b), (d), (f), and (h), are corrected using the iterative approach described in the text.
    These data have an average accuracy of 33~\unit{\milli \radian} in $\theta$, and 43~\unit{\milli\radian} in $\Phi$.
    The black lines with a slope of one are included to guide the eye.
    }
    \label{fig:measured_angles}
\end{figure*}

To assess this approach to measuring the signal $k$-vector, we rotate the signal horn around the cell, sweeping over a grid of polar angles from $\pi/4$ to $3\pi/4$ and azimuthal angles from $\pi/8$ to $3\pi/8$ with a step size of $\pi/36$ (5$^\circ$) representing a solid angle of $\pi/2\sqrt{2}$.
We also sweep the phase offset $\phi_{vh}$ over $2\pi$ to apply fields with different ellipticities.

At each $k$-vector, we model variations from the expected behavior due to rf reflections present in the interaction region.
This modeling, which also includes corrections for the LO amplitudes and phases, is previously described in Ref.~\cite{Elgee24}, and provides a matrix ($A$) that represents the amplitudes and phases of the reflections from one axis into another.
For instance, $A_{ij}$ represents the amplitude and phase of LO$_{i}$ along axis $\hat{j}$.
As the signal horn moves to different $k$-vectors, the character of the rf reflections changes.
We first ignore this variation and apply the average correction equivalently across all $k$-vectors, which works mostly to calibrate the LO amplitudes and phases.
The average correction parameter matrix $A$ is
\begin{equation}
A = 
    \begin{bmatrix}
        1.003 & 0.028e^{1.75} & 0.034e^{0.20}\\
        0.050e^{0.24} & 0.948e^{-2.10} & 0.020e^{0.98}\\
        0.003e^{0.88} & 0.017e^{-3.13} & 1.000e^{0.47}
    \end{bmatrix}
\end{equation}
where the diagonal elements primarily represent corrections to the LO amplitudes and phases, and the off-diagonal elements represent reflection amplitudes and phases.
The resulting polar and azimuthal angles from this averaged correction, restricted to purely left or right circularly polarized input fields, are presented in Fig.~\ref{fig:measured_angles} (a), (c), (e), and (g).
The accuracy of this method, defined as the average difference between the applied and measured angles, is 60~\unit{\milli\radian} in $\theta$ and 63~\unit{\milli\radian} in $\Phi$.
In these data we see a systematic coupling between $\theta$ and $\Phi$.
Most clearly, for left circularly polarized light $\Phi$ is underestimated at low $\theta$.
This effect is likely due to the changing character of the rf reflections as a function of $k$-vector.

Over the measured $k$-vectors, the elements of the parameter matrix $A$ vary with standard deviations in amplitude ($\sigma_{|A|}$) and phase ($\sigma_\phi$) respectively of 
\begin{equation}
\sigma_{|A|} = 
    \begin{bmatrix}
        0.144 & 0.036 & 0.036\\
        0.027 & 0.097 & 0.033\\
        0.033 & 0.038 & 0.119
    \end{bmatrix}, \ 
\sigma_\phi =     \begin{bmatrix}
        0 & 1.20 & 0.99\\
        0.48 & 0.09 & 1.12\\
        1.85 & 2.23 & 0.13
    \end{bmatrix}.
\end{equation}
We account for these changing reflections and address the systematic coupling between $\theta$ and $\Phi$ by applying an iterative correction approach.
First, we create a map of reflection parameters as a function of the true input $k$-vector.
Then, for a given data point, we take the measured $k$-vector using the average method described above as a best guess, and apply the reflection parameters from the closest $k$-vector in the map.
This generates a new and more accurate $k$-vector, and the process converges rapidly when iterated.
The measured angles found using this method with three iterations are shown in Fig.~\ref{fig:measured_angles} (b), (d), (f), and (h).
In these data, the systematic coupling between $\theta$ and $\Phi$ is largely removed, and the accuracy is improved to $33$~\unit{\milli\radian} in $\theta$, and $43$~\unit{\milli\radian} in $\Phi$.
The remaining deviations are likely from instabilities in the iterative correction where the initially-measured $k$-vector is too far off from the true value, or the structure in the rf reflections is too fine to be captured by the map.

{\it Interaction region averaging}---Our measurement averages over the spatial extent of the interaction region.
When the $k$-vector of the signal field is not aligned with the $k$-vector of a particular LO, and the interaction region is an appreciable fraction of the carrier wavelength, the phase of the respective heterodyne beat will vary in space.
Our measured signal is the result of averaging over this phase gradient.
In our system, the optical beams are negligibly small relative to the carrier wavelength with a ratio of $\lesssim 0.02$, so only the averaging along the optical axis ($\hat{z}$) is significant.
Along this axis, the phase gradient of the beat has a wavelength ($\lambda_p$) given by
\begin{equation}
    \lambda_p = \lambda_0/|\cos(\theta)|
\end{equation}
where $\lambda_0$ is the wavelength of the rf carrier.
When we average over the interaction region of length $d$ we get a normalized amplitude reduction of 
\begin{equation}\label{eq:interaction_averaging}
    |b_i|/|\text{Sig}_i| \propto |\text{sinc}(\pi d/\lambda_p)|,
\end{equation}
where Sig$_i$ is the predicted amplitude of the signal field projected onto axis $\hat{i}$.
Since, in our system, all LO's are perpendicular to the optical axis, all the beat amplitudes should average equally.
This should only affect the measured total amplitude of the field and the noise in the $k$-vector measurement, but not the accuracy.

\begin{figure}[b]
    \includegraphics[width = 3.375 in]{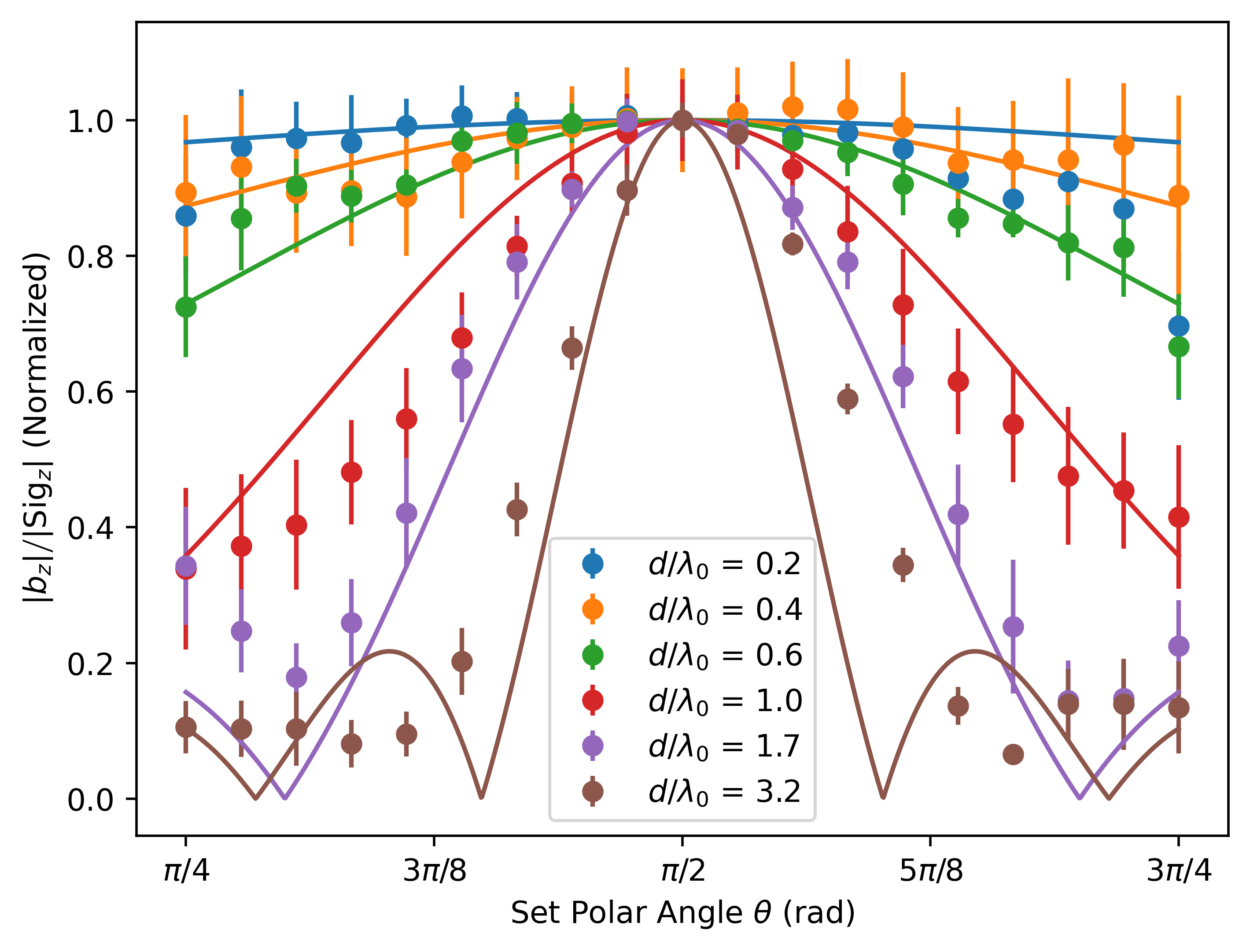}
    \caption{The effect of averaging over the interaction region on the ratio of the measured beat amplitude $|b_z|$ to the expected $z$ signal projection $\text{Sig}_z$ as a function of $\theta$ for a variety of cell lengths and carrier wavelengths.
    The solid lines correspond to the predicted effect from Eq.~\ref{eq:interaction_averaging}.}
    \label{fig:region_averaging}
\end{figure}

We evaluate this averaging effect by testing the system with three different vapor cell lengths (1~\unit{\centi\meter}, 2.5~\unit{\centi\meter}, and 7.6~\unit{\centi\meter}), and two different carrier wavelengths (4.5~\unit{\centi\meter} and 2.4~\unit{\centi\meter}).
The carrier wavelengths correspond to the 70D$_{5/2} \rightarrow 69$F$_{7/2}$ resonance at 6.64~\unit{\giga\hertz} and the 57D$_{5/2} \rightarrow 56$F$_{7/2}$ resonance at 12.47~\unit{\giga\hertz}.
These parameters yield values for $d/\lambda_p$ ranging from 0.2 to 3.2.
The result of this averaging effect on $|b_z|/|\text{Sig}_z|$, normalized to the value at $\theta = \pi/2$, can be seen in Fig.~\ref{fig:region_averaging}.
The data exhibit the expected behavior, with deviations likely due to changes in the absorption and reflection of rf through the cell at different angles.
The minima of the $\text{sinc}$ function correspond to locations where the signal is averaged over exactly a single period of the phase gradient, and are resolvable in the data for $d/\lambda_0 = 1.7$ where this constraint is satisfied at $\theta = \pi/2 \pm 0.63$.
Although not shown here, the averaging effect looks similar in $|b_x|/|\text{Sig}_x|$ and $|b_y|/|\text{Sig}_y|$ for the longer wavelength $\lambda_0 = 4.5~\unit{\centi\meter}$.
However, when $\lambda_0 = 2.4~\unit{\centi\meter}$, close to the cell diameter of $2.54$~\unit{\centi\meter}, and the signal is normal to the cylindrical walls of the cell, $|b_x|/|\text{Sig}_x|$ and $|b_y|/|\text{Sig}_y|$ vary significantly from theory, and from cell to cell. 
We attribute this to interference effects from reflections off the cell walls, providing evidence that most of the reflection effects we see in the system are due to the vapor cell rather than the external environment.

These data confirm that the solid angle over which the $k$-vector can be measured increases as the interaction region becomes small relative to the carrier wavelength.
In general, there will be a trade-off between absolute sensitivity, which typically improves with interaction volume through increased atom number, and the sensor's field a view.
However, in principle the sensor can be made arbitrarily electrically small.
For instance, we are able to measure the $k$-vector in three dimensions even when the sensor is significantly electrically small along $\hat{x}$ and $\hat{y}$.

\begin{figure}[b]
    \includegraphics[width = 3.375 in]{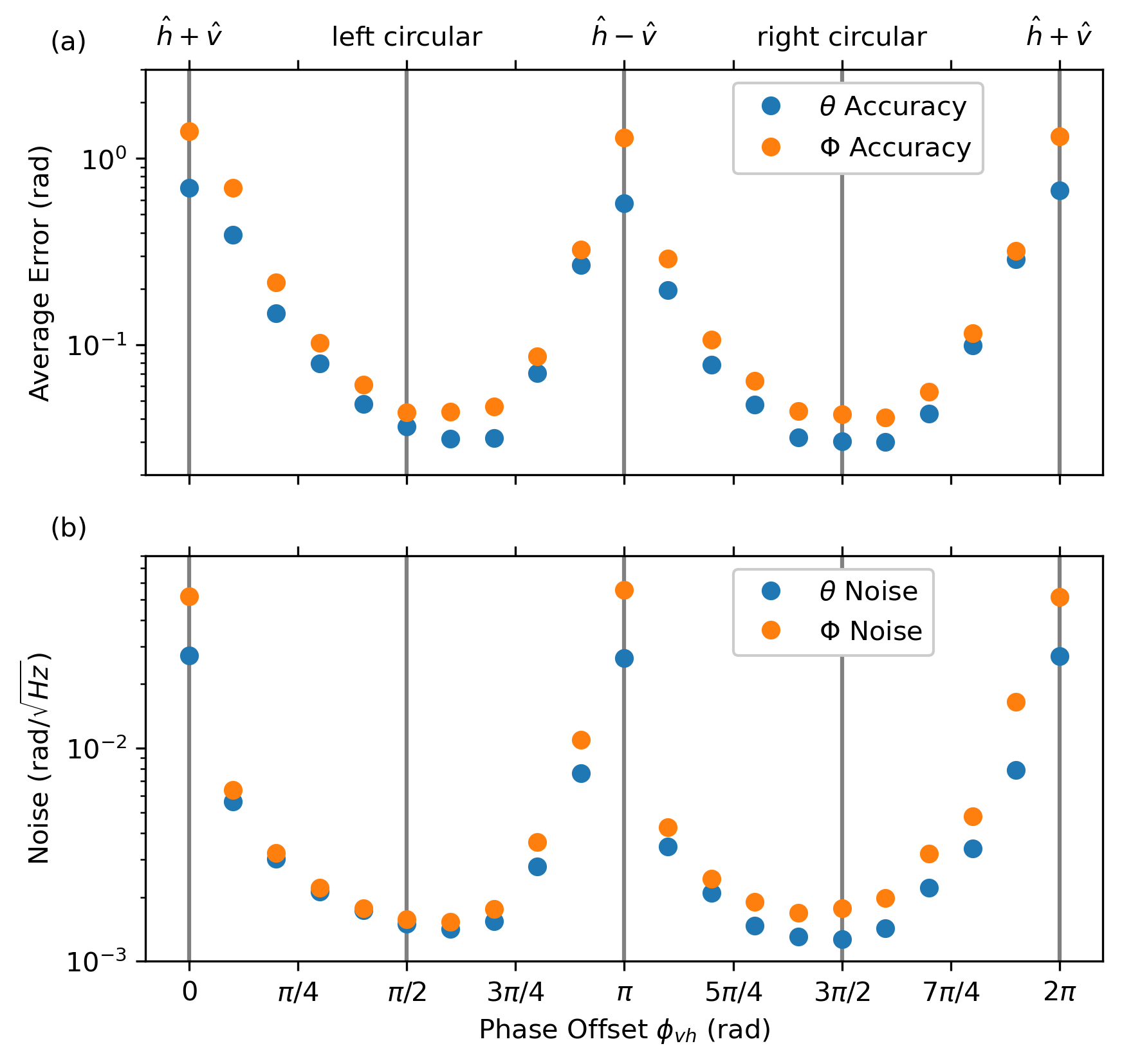}
    \caption{
    The accuracy (a) and noise (b) in the measured polar and azimuthal angles as a function of the polarization ellipticity.
    }
    \label{fig:ellipticity_error}
\end{figure}

{\it Ellipticity dependence}---For circular or elliptical signal polarization one can define a single $k$-vector perpendicular to the electric field at all times.
For pure linear polarization the $k$-vector can only be restricted to a plane.
Thus, our ability to extract the $k$-vector depends on the ellipticity of the input field.
To investigate this, we scan the phase offset between the $\text{Sig}_{v}$ and $\text{Sig}_{h}$ inputs, which changes the signal between linear and circular polarization.
Fig.~\ref{fig:ellipticity_error} (a) shows the average magnitude of the error in the measured $\theta$ and $\Phi$ as a function of the signal phase offset $\phi_{vh}$ averaged over all measured $k$-vectors.
These data use the iterative correction method described above.
The error exhibits the expected sharp peaks near linear polarization, but is flat near circular polarization.
The values at pure linear polarization are likely due to a combination of residual ellipticity, and the error associated with restricting the $k$-vector to a plane.
There is some asymmetry which most notably causes the minimal error in $\theta$ to be pushed off from perfectly circular light, and which we attribute to residual effects from reflections.

We also investigate the statistical error per $k$-vector by splitting the data trace of each measurement into 10 time bins, each 10~\unit{\milli\second} long, extracting the standard deviation in the $\theta$ and $\Phi$ measurements, and normalizing to the integration time.
The average noise in the angle measurements as a function of signal phase offset is shown in Fig.~\ref{fig:ellipticity_error} (b).
Compared to the accuracy, the noise has a slightly flatter dependence on signal ellipticity, but the asymmetries remain.
The minimal noise is 1.3~\unit{\milli\radian\per\sqrt{\hertz}} for $\theta$ and 1.5~\unit{\milli\radian\per\sqrt{\hertz}} for $\Phi$.
The difference in the noise for $\theta$ and $\Phi$ is expected as $\Phi$ involves more measured quantities in its calculation.
The statistical noise here corresponds to a field sensitivity of 58~\unit{\micro\volt\per\meter\sqrt{\hertz}}, which is similar to the sensitivity of the previous experimental iteration \cite{Elgee24}.
In all, these results show that the $k$-vector can still be extracted over a wide range of ellipticities.

{\it Conclusions}---We have demonstrated $k$-vector measurements with an electrically-small Rydberg electric field sensor for elliptically-polarized input fields.
The accuracy of the current sensor is likely limited by rf reflections inside the vapor cell.
Vapor cell design, or more precise modeling of reflections, could lessen this effect.
Additional development could also address the trade-off between a small interaction region and signal size.
For instance, our method should be compatible with a non-collinear three-photon Doppler-free excitation scheme \cite{Ryabtsev11}, which could provide higher signal sizes for small interaction regions.
The inability of the sensor to measure the $k$-vector of linearly polarized fields could also be addressed by incorporating rf phase plates to circularize the input field \cite{Hernandez19, Rohrbach21, Jackel22}.
However, in some cases the birefringence of the propagation medium, such as the ionosphere, may provide sufficiently elliptical signals \cite{murphy96, Carrano23}.

Our sensor employs the power of rf heterodyne measurements in Rydberg atom electric field sensors to extract the $k$-vector in three dimensions with a single measurement in a platform that can be arbitrarily electrically small.
This expands the potential applications of these sensors and demonstrates a more full characterization of the measured field than previously possible with this platform. 

{\it Acknowledgments}---We thank Fredrik Fatemi for valuable discussions.
The views, opinions and/or findings expressed are those of the authors and should not be interpreted as representing the official views or policies of the Department of Defense or the U.S. Government.

\appendix

\bibliography{k-vector}

\end{document}